\documentstyle[12pt,aasms4]{article}
\newcommand{\etal}{{\it et al.} }
\newcommand{\asca}{{\it ASCA} }

\newcommand{\rosat}{{\it ROSAT} }
\newcommand{\mcg}{MCG $-$6$-$30$-$15 }
\newcommand{\pks}{PKS~0637$-$752 }
\newcommand{\psrc}{PKS~2149$-$306 }
\newcommand{\cosmo}{$H_{0} = 50 \rm \ km \ s^{-1} \ Mpc^{-1}$ and
$q_{0}=0$}

\begin{document}

\title{A HIGHLY DOPPLER BLUESHIFTED Fe-K 
EMISSION LINE IN THE HIGH-REDSHIFT QSO \psrc}

\author{T. Yaqoob\altaffilmark{1,2},
I. M. George, \altaffilmark{1,2},
K. Nandra\altaffilmark{1,2},
T. J. Turner\altaffilmark{1,3},
S. Zobair\altaffilmark{1},
P. J. Serlemitsos\altaffilmark{1}}

\altaffiltext{1}{Laboratory for High Energy Astrophysics, 
NASA/Goddard Space Flight Center, Greenbelt, MD 20771, USA.}
\altaffiltext{2}{Universities Space Research Association}
\altaffiltext{3}{University of Maryland, Baltimore County.}

\vspace{5cm}

\begin{center}
{\it Accepted 2 September 1999 for publication in Astrophysical Journal Letters}
\end{center}

\begin{abstract}
We report the results from an \asca observation of the 
high-luminosity, radio-loud quasar \psrc (redshift 2.345), covering the 
$\sim 1.7-30$~keV band in the quasar-frame.
We find the source to have a 
luminosity $\sim 6 \times 10^{47}\ {\rm ergs \ s^{-1}}$ in the
2--10~keV band (quasar frame).
We detect an emission line centered at $\sim 17$ keV in the
quasar frame.
Line emission at this energy has not been observed in any
other active galaxy or quasar to date. 
We present evidence rejecting the possibility that this line is 
the result of instrumental artifacts, or a
serendipitous source.
The most likely explanation is blueshifted Fe-K emission
(the equivalent width, is $\rm EW  \sim 300 \pm 200$ eV, quasar frame).
Bulk velocities of the order of $0.75c$ are implied by the data.
We show that Fe-K line photons originating in an accretion disk
and Compton-scattering off a leptonic jet aligned along the disk
axis can account for the emission line.  
Curiously, if the emission-line feature recently
discovered in another quasar (PKS 0637$-$752, $z=0.654$) at 1.6 keV in the
quasar frame, is due to blueshifted O{\sc vii} emission, the
Doppler blueshifting factor in both quasars is similar ($\sim 2.7-2.8$).
\end{abstract}

\section{INTRODUCTION}

X-ray emission lines are powerful diagnostics of the physical
conditions and structure in active galactic nuclei (AGNs).
The picture emerging from many X-ray studies of AGNs over
the last couple of decades 
is that emission
features in the X-ray spectra
become scarce in AGNs
with 2--10 keV intrinsic luminosity exceeding $\sim 10^{45}$
$ \rm ergs \ s^{-1}$. This trend 
is discussed at length in Nandra \etal (1996, 1997a) and Reeves \etal (1997;
and references therein). 
The transition to a featureless
X-ray power-law continuum 
in the vast majority of the high luminosity AGNs
is not fully understood, but may be related to the complete
ionization of matter responsible for emission-line features and/or
beaming of the X-ray continuum swamping out emission-line features.
Any emission features observed in type 1 AGN 
are almost always Fe-K lines, although there are a handful of reports
of other features (George, Turner, and Netzer 1995;
Fiore \etal 1998; 
Turner, George and Nandra 1998), all
in low-luminosity AGNs.
However, Yaqoob \etal (1998) reported an 
emission line in the high-luminosity radio-loud quasar 
PKS~0637$-$75 ($z=0.654$, $L[2-10 \rm \ keV] \sim 10^{46} \ ergs \ s^{-1}$). The line energy was 1.6 keV in the rest-frame and was interpreted
as highly blueshifted O{\sc vii}.

In this paper we report another X-ray emission-line
feature at an unexpected energy, 
in the radio-loud quasar \psrc ($z=2.345$). With
$L[2-10 \rm \ keV] > 10^{47} \ ergs \ s^{-1}$, this makes \psrc
the most luminous and highest redshift AGN in which an X-ray emission
line detection has been claimed. 
The \asca ({\it Advanced Satellite for Cosmology and Astrophysics};
Tanaka, Inoue and Holt 1994) observation of \psrc 
reported here has also been studied by Siebert \etal (1996) and 
Cappi \etal (1997). The emission line is 
visible by eye in their work, but went unnoticed because it
is common practice to only search for emission lines at energies
one would expect.  

\section{THE \asca DATA}

\asca observed \psrc on 1994 October 26--27 for a duration
of $\sim 50$ ks. 
The reader is referred to Tanaka \etal (1994)
for details of the
instrumentation aboard {\it ASCA}. The two Solid State
Imaging Spectrometers (SIS),
hereafter SIS0 and SIS1, with
a bandpass of $\sim 0.5-10$ keV, were operated in 1-CCD FAINT and BRIGHT
modes. The two Gas Imaging Scintillators (GIS),
hereafter GIS2 and GIS3, with a bandpass
of $\sim 0.7-10$ keV, were operated in
standard PH mode. 
The data reduction procedures follow exactly those described
in Yaqoob \etal (1998) and the reader is referred to that paper for
the details of the data reduction, relevant calibration, 
other systematic uncertainties, and screening criteria. 
Screening resulted in net `good' exposure times in the range $\sim 
19.4$--19.9 ks for the four instruments.

Images were accumulated from the
screened data and they
were examined for possible nearby contaminating
sources. No such sources were detected.
Source events were extracted from circular regions with radii
of 4' for the SIS and 6' for the GIS. Background events were 
extracted from off-source regions. As a check, background spectra
were also made from a 1-CCD mode blank-sky observation which resulted
from the non-detection of the quasar IRAS 15307+3252 (observed 1994
July). These background data were subject to identical selection criteria
and used to check the invariance of the spectral results for
\psrc described below.
The count rates in the accumulated background-subtracted
spectra for \psrc ranged from $\sim 0.17$--0.30 ct/s. Spectra were binned
to have a minimum of 20 counts per bin in order to utilize
$\chi^{2}$ as the fit statistic.
  
The \asca lightcurves do not show significant variability,
the highest excess variance (e.g. see Nandra \etal 1997b) 
we obtain is from SIS0 and has the value
$(6.9 \pm 3.7) \times 10^{-3}$ for 128 s bins.
This is to be compared
with  $0.111 \pm 0.005$ we obtain 
for the highly variable Seyfert 1 \mcg from a 4 day \asca
observation.

\section{SPECTRAL FITTING RESULTS}

We fitted spectra from the four \asca instruments simultaneously
in the range 0.5--10 keV with a simple power-law plus 
cold, neutral absorber model. 
A total of two interesting parameters were involved (the
photon index,  $\Gamma$, and column density, $N_{H}$), plus
four independent instrument normalizations. (The deviation
of any of the four normalizations from their mean was less than 7\%).
The results are shown in Table 1.
The best-fitting photon index,  $\Gamma = 1.54 \pm 0.05$, is consistent
with that reported by Siebert \etal (1996) and Cappi \etal (1997)
and is rather flat. 
The column density, $N_{H}$, is at least a factor 3 higher than the Galactic
value of $2.12 \times 10^{20} \ \rm cm^{-2}$ obtained from
Dickey and Lockman (1990). 
The instrument-averaged 
{\it observed} 0.5-2 keV and 2--10 keV fluxes are
3.6 and $10.1 \times 10^{-12} \rm \ ergs \ cm^{-2} \ s^{-1}$ 
respectively. The 2--10 keV luminosity in the source frame
is $5.8 \times 10^{47} \rm \ ergs \ s^{-1}$ (\cosmo ). 

Although the above simple continuum
model provides an adequate fit, inspection of the ratios of data to
model (Figure~1) reveal a statistically significant 
(see below) `hump' at $\sim 5$ keV
(observed frame) in both SIS0 and SIS1. 

We repeated the four-instrument \asca fits
with the addition of a simple Gaussian to model the apparent emission-line
feature. Three extra parameters were
involved, namely the line center energy, $E_{c}$; the
Gaussian intrinsic width, $\sigma$, and the line
intensity, $I$. {\it Note that we will refer to all three
parameters, plus the equivalent width (EW), in the quasar frame}.
Initially we allowed all three parameters to float 
but the line width was not constrained (the best fit
giving $\sigma \sim 1$ keV with only $\Delta \chi^{2} = 1.2$
relative to the case for a narrow line).
The results for a narrow-line fit with $\sigma$ fixed at 0.01 keV
are given in Table 1.
Note that the decrease in $\chi^{2}$ compared to the
model without an emission line is $10.3$ for the addition of
two free parameters so the line feature is detected at
a confidence level greater than 99\%. It is appropriate here to use 
a $\Delta \chi^{2} = 4.61$ criterion to derive 90\% confidence errors
on the best-fitting line parameters since we want to know the bounds
on $E_{c}$ and EW, independently of $\Gamma$ and $N_{H}$ (e.g. see Yaqoob 1998).
We obtain,
approximately, $E_{c} = 17.0 \pm  0.5$ keV and $\rm EW = 300 \pm 200$ eV. 
We also tried fitting SIS0 and SIS1 with a narrow line separately
and obtained $\Delta \chi^{2} = 5.1$ and 6.8 respectively. From Figure 1
it appears that $\Delta \chi^{2}$ should be higher for SIS0, not SIS1.
However, these results are for a {\it narrow line}. The SIS0 data,
which have the better statistical quality, prefer a somewhat broader 
feature. We also fitted a narrow line to the GIS2 plus GIS3 data only,
this time with the energy fixed at 17 keV, and obtained a 90\% 
($\Delta \chi^{2}=2.7$) upper limit on the intensity of
$6.2 \times 10^{-5} \rm \ photons \ cm^{-2} \ s^{-1}$, consistent
with the intensity in Table 1.
The feature is not obvious in the GIS; this is
not surprising and it is quite common for
weak Fe-K lines in AGN to be detected in the SIS but not the GIS.
The GIS count rates are a factor 1.5--2 lower than the SIS
(due to larger off-axis angles) and the energy resolution is
a factor $\sim 4$ times worse. 
We also examined public archive data from a {\it BeppoSAX} observation
of \psrc performed in October 1997. No line was detected but the data
do not rule it out. Again, the count rates (in the MECS) were much
lower than SIS (a factor of $\sim 6$ less) and the energy resolution
worse by a factor of $\sim 4$.

\section{THE REALITY OF THE EMISSION-LINE FEATURE}

Since an emission-line feature like that reported
above has never before been observed in any AGN,
we sought explanations which did not
require the line to be intrinsic to the quasar. 

Firstly, is it possible that some or all of the X-ray emission is
due to another source? \psrc is in the \rosat All-Sky
Survey Bright Source Catalog but no other X-ray source is found within
11 arcmin. However, the Cambridge APM shows up an extended
source only 2 arcmin away, of similar optical magnitude. This source was
not found in either the SIMBAD or NED databases and remains a mystery.
In case this latter source is absorbed but still contributing to the
hard X-rays, we compared the PSF of the \asca \psrc data with the 
PSF for a bright
point source (3C 273), and also with the PSF of an
extended source (the cluster E0657$-$56, see Yaqoob 1999).
The results are shown in Figure 2.
The \asca data are fully consistent
with the X-ray emission from a single point source corresponding to 
PKS 2149$-$306.
We also note the bright cluster
$\sim 40$ arcmin away, Abell 3814, cannot be responsible for
contamination of the \asca data either. Besides, its redshift is 0.118 
so the Fe-K line complex would be observed around 6 keV, whereas we
observe  
the emission line in \psrc at $\sim 5.1$
keV.

To demonstrate that the SIS calibration is not responsible we examined
SIS data from an observation of the X-ray binary
4U 1957$+$11, performed $\sim 4$ days after the \psrc observation.
Figure 3 shows the ratios of data to  a
simple power-law plus absorber model
(both SIS0 and SIS1) 
fitted over the 2--10 keV range.
It can be seen that no emission line feature at 5.1 keV
is present in these data. We also made new SIS spectra for \psrc 
artificially changing the gain by $\pm 1\%$. This did not change the
results - the emission line was still present, with insignificant 
effect on the line parameters. There is the remaining possibility
that some other, unknown instrumental artifact, occurring only during
the PKS 2149 observation, was responsible for producing or
mimicking an emission line. We consider this highly unlikely, however,
particularly as the line is clearly detected in both SIS.

The emission line cannot be a
background feature (for example from Galactic diffuse emission)
because the background from nearby regions in the same
field and at the observed energy of the emission line in the SIS
lies at least a  factor 20 below the on-source spectrum.
Also, we repeated the spectral analysis using the alternative
background spectra made from a blank field, as described in \S 2, and 
the results were confirmed to be robust to the background used. 
It is unlikely that the emission line is produced in an intervening
galaxy 
since that 
galaxy would have to be extremely bright
for the emission line to be detectable above
the quasar continuum, with a redshift of
$\sim 0.3$. Such a galaxy would be bright enough that
signatures
in the optical would have been detected.

Finally, we can ask, what is the chance probability that a line-like feature
is detected in {\it both} SIS0 and SIS1, due to random statistical
fluctuations, if we examine a sample of high-redshift quasars?
Suppose the quasar spectra have $N$ energy bins.
The probability of
finding a {\it positive} fluctuation,
relative to some continuum model, of more than
$2 \sigma$ (i.e. about the level of detection in {\it each}
SIS for \psrc), in one or more energy bins, in SIS0 say,
is $1- [1-p]^N$ where
$p=\int_{+2\sigma}^{\infty} \exp{(-x^{2}/2\sigma^{2})} \ dx =
2.276 \times 10^{-2}$ (assuming a Gaussian distribution). Then the probability
that the same feature is detected at more than $2 \sigma$ in SIS1, in
the same energy bin is $P=p(1-[1-p)]^N)$.
Note that the $\sim 2\sigma$ detection in {\it each} SIS is
consistent with the $\sim 3\sigma$ net detection as indicated by the
$\Delta \chi^{2}$ from the spectral fits (see Table 1).
We examined all quasars with $z>0.158$, for which the \asca data
were public as of 1998, December 4. This amounted to
87 observations of 86 sources. Of these, only 31 datasets yielded
spectra which were of sufficient quality to search for line-emission
(specifically, we defined this so that the SIS0 spectrum had to
be brighter than 0.05 ct/s). Assuming the highest value for
the number of bins in each SIS spectrum ($N=155$) in order
to get the most conservative estimate, we obtain $P=0.022$.
Thus, of the 31 datasets we expect less than one (0.69) to
yield an emission-line in both SIS0 and SIS1 from
random fluctuations alone. Then, for the 31 spectra,
we measured the $\Delta \chi^{2}$
resulting from inserting a narrow Gaussian emission line
at 100 energies between 0.5 and 8 keV for each dataset,
relative to a simple power-law
plus Galactic absorption continuum model, fitted simultaneously to
four instruments.
We found four sources for which $\Delta \chi^{2}>9$ (i.e. a detection
significance of $>3 \sigma$ for 1 interesting parameter).
These were E1821$+$143 (see Yamashita \etal 1997),
PG 1116$+$265 (see Nandra \etal 1996),
\pks (see Yaqoob \etal 1998), and PKS 2149$-$306.
The probability of finding four such sources by chance is
$(31!/27!/4!) P^4 (1-P)^{27}$, or 0.4\%. Even without the two sources
with highly probable bona-fide Fe-K lines,
the probability is $(29!/27!/2!) P^2 (1-P)^{27}$, or 10.9\%.

\section{DISCUSSION}

The most likely explanation for the detection of an emission line centered
at $\sim 17$ keV in the rest frame of the radio-loud quasar
PKS~2149$-$306 ($z=2.345$) is blueshifted Fe-K emission
(at a rest-energy in the range 6.4--6.97 keV).
The Doppler factor (ratio of Lorentz-boosted
energy to rest-energy), $D= 2.42-2.65$
implies bulk-motion velocities of $ 0.71c -0.75c$
(head-on). 
Interpreting the emission line in \psrc as highly blueshifted
Fe-K carries with it some strong physical constraints with respect
to its origin, in order to account for the large rest-frame
equivalent width (EW) of ($300 \pm 200$ eV). 

The main problem is for the line-emitting material to present
enough solid angle to the X-ray continuum source to produce
sufficient line equivalent width, but not so much that the
line becomes too broad. These conflicting requirements must
be overcome by any model.
For example, although a spherical
distribution of optically-thick clouds 
(e.g. see Nandra and George 1994) could produce the necessary
equivalent width with only a modestly small covering factor
of $\sim 3\%$ and infall with $\beta \equiv v/c =0.75$, the fact that
one must integrate over all line-of-sight angles would give a
broad line profile stretching from  $\sim 2.4-17$ keV in the
rest frame. (Infall, as opposed to outflow or a chaotic velocity
field gives the largest boost to the equivalent width 
since the continuum as seen by the Fe atoms is 
maximally Lorentz-boosted, as well as the line emission).
 
Here we propose a possible model which overcomes the
conflicting solid-angle requirements.
We suppose that an Fe-K emission line is produced in an accretion disk,
just as thought to be the case for Seyfert 1 galaxies and some quasars
with redshift lower than that of PKS 2149$-$306. Further, the line
is assumed to be narrow (relative to the \asca 
spectral resolution of $\sim 150$ eV at Fe-K), 
like that observed in the high-luminosity quasar 
PG 1116$+$205 (i.e. not as broad as the Fe-K lines observed in
Seyfert 1). 
Then, we
invoke a leptonic jet aligned along the disk axis, with the leptons
outflowing at $\beta \sim 0.75$. For the sake of argument,
it is assumed that the observer's line of sight is aligned
with the jet axis (it doesn't have to be exactly aligned).
If the jet is very long compared to the
disk radius then the requirement of a small solid angle is satisfied,
(resulting in a line-profile which is not too broad,
especially in view of the tightly constrained velocity field
of the jet), and yet can
potentially intercept most  of the Fe-K line photons from the disk
which escape parallel to the disk and jet axis. 
The disk photons will be Compton-scattered by the leptons in the
jet and will be highly collimated along the jet axis since
the line photons leaving the disk and intercepting the jet
are strongly aligned 
along the disk/jet axis
and the Doppler boosting by the jet favours forward scattering, the
larger the Lorentz factor. This again ensures that the line does not become
too broad. Now, both line and continuum photons will
experience the same Doppler effects so the only factor
contributing to an increase in the equivalent width (EW) of the line
is the fact that the latter is defined as  
ratio of line to continuum {\it per unit energy} so the
intrinsic EW ($\equiv \rm EW_{0}$) 
will be increased by a factor $F \equiv Df[1-\exp{(-\tau})]$ where
$f$ is the fraction of the disk area blocked by the jet and
$\tau$ is the jet Thomson optical depth. 
Actually the optical depth
will vary  at different points along the jet so $\tau$ represents
the value averaged over all the interactions of the disk line-photons
with the jet. In principle, $\rm EW_{0}$ could be as large $\sim 300$ eV
(as observed in Seyfert galaxies). This would imply 
$F  \sim 0.4$;
on the other hand if $\rm EW_{0}$ is a few tens of eV (say, 50 eV), 
$F \sim 2$.  
Multiple scattering is not a problem because the velocity field
is so narrow that multiply-scattered photons will appear as narrow
(weaker) features at higher energies. 
Such a fine-tuning of $\tau$ also may explain why
the phenomenon is not commonly observed in other quasars. 
This model also predicts that nothing unusual would be observed in the
optical emission-line spectrum 
(and indeed this is the case, e.g. Wilkes 1986)
if the BLR is situated much further
from the central X-ray source than the disk-jet system. 
 
We thank Andy Fabian and Rick Edelson for discussions on the results presented
in this paper. We also thank an anonymous referee for helping to
improve the paper.
We thank the \asca mission operations team at ISAS, Japan,
and all the instrument teams 
for their dedication and hard work in making these \asca observations 
possible. This research made use of the HEASARC archives at the
Laboratory for High Energy Astrophysics, NASA/GSFC and
the NASA/IPAC Extragalactic Database (NED) which is operated by JPL,
CALTECH, under contract with NASA.

\newpage

\begin{deluxetable}{lccc}
\tablecaption{Spectral Fits to \psrc}
\tablecolumns{4}
\tablewidth{0pt}
\tablehead{
\colhead{Parameter} & \colhead{No line} &
                \colhead{Narrow line}
}

\startdata

$\Gamma$
        & $1.54^{+0.05}_{-0.05}$ 
        & $1.56^{+0.05}_{-0.05}$ \nl
$N_{H}$ ($10^{20} \ \rm cm^{-2}$)
        & $7.0^{+2.3}_{-2.3}$   
        & $7.5^{+2.4}_{-2.3}$  \nl
$\sigma$ (keV)
        &                             
        & 0.01 (FIXED) \nl
$E_{c}$ (keV)
        &                            
        & $17.00^{+0.52}_{-0.48}$ \nl
$I$
        &                    
        & $4.5^{+3.1}_{-3.1}$ \nl
EW (eV)
        &                   
        & $298_{-205}^{+202}$ \nl
$\chi^{2}$
        & 585.6 
        & 575.3  \nl
degrees of freedom
        & 578 
        & 576 \nl
\tablecomments{The continuum model used is a simple power law
plus absorber with
the photon index, $\Gamma$, and column density, $N_{H}$, floating.
The absorber is placed at $z=0$. The emission line parameters
are all referred to in the quasar frame ($z=2.345$); these
are the intrinsic width, $\sigma$, the center energy, $E_{c}$,
the intensity ($\rm 10^{-5} \rm \  photons \  cm^{-2} \ s^{-1} $),
and the equivalent width (EW).
Errors are 90\% confidence for two interesting parameters
($\Delta \chi^{2} = 4.61$).}
\enddata
\end{deluxetable}

\newpage
\section*{Figure Captions}

\par\noindent
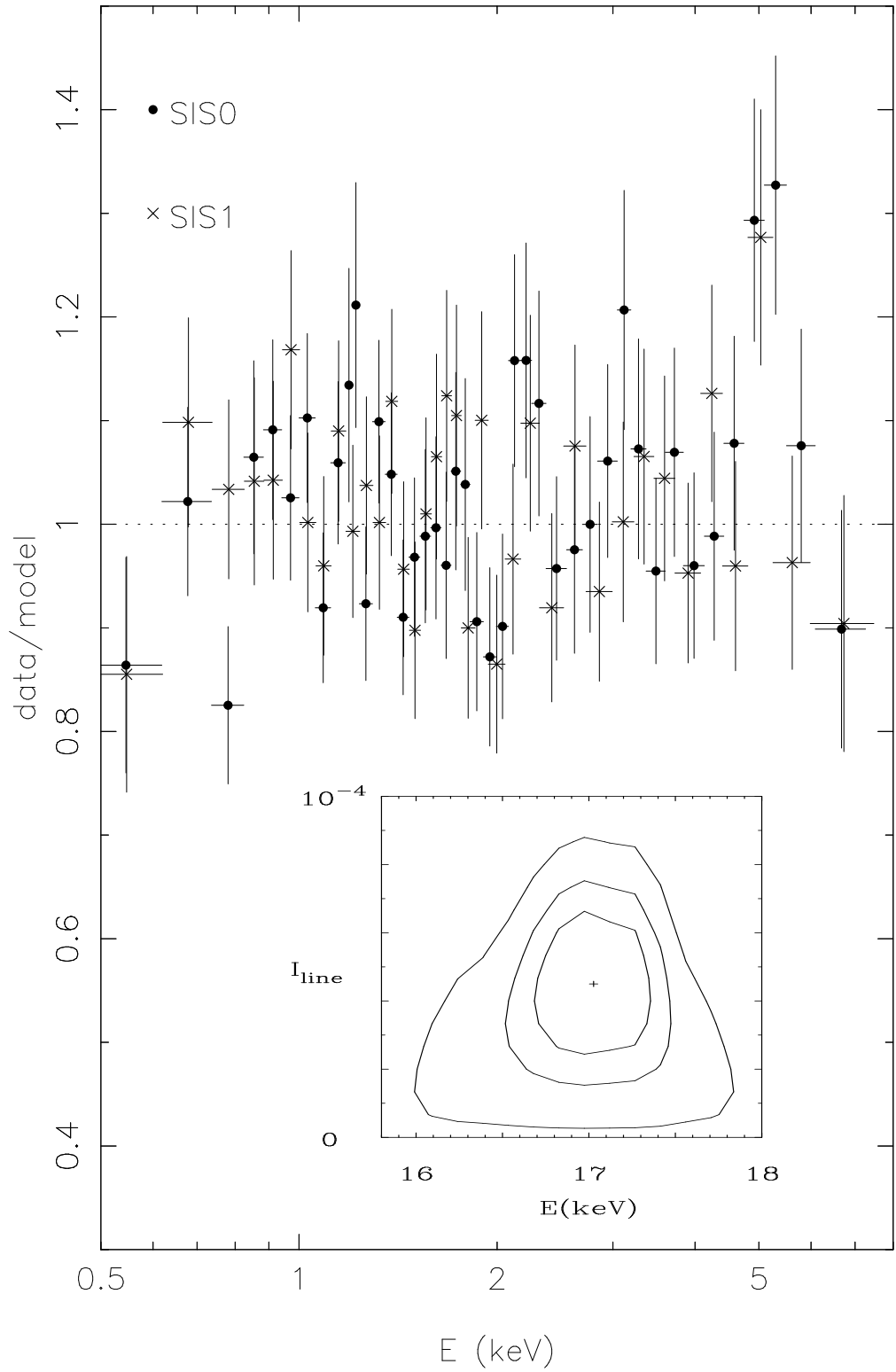
{\bf Figure 1} \\
Ratio of the \asca SIS spectral data for \psrc ($z=2.345$)
to the best-fitting power-law plus
absorber model versus {\it observed} energy. 
Notice the significant residuals at $\sim 5$ keV, 
corresponding to an emission line centered at
$\sim 17$ keV in the quasar frame (see text).
The inset shows 68\%, 90\% and 99\% confidence contours 
versus {\it rest-frame} energy, which
result from fitting a Gaussian emission line to the residuals,
corresponding to $\Delta \chi^{2} = 2.28, 4.61$, and 9.21
respectively. The line intensity, $I$, is in units of
$\rm photons \ cm^{-2} \ s^{-1}$. 

\par\noindent
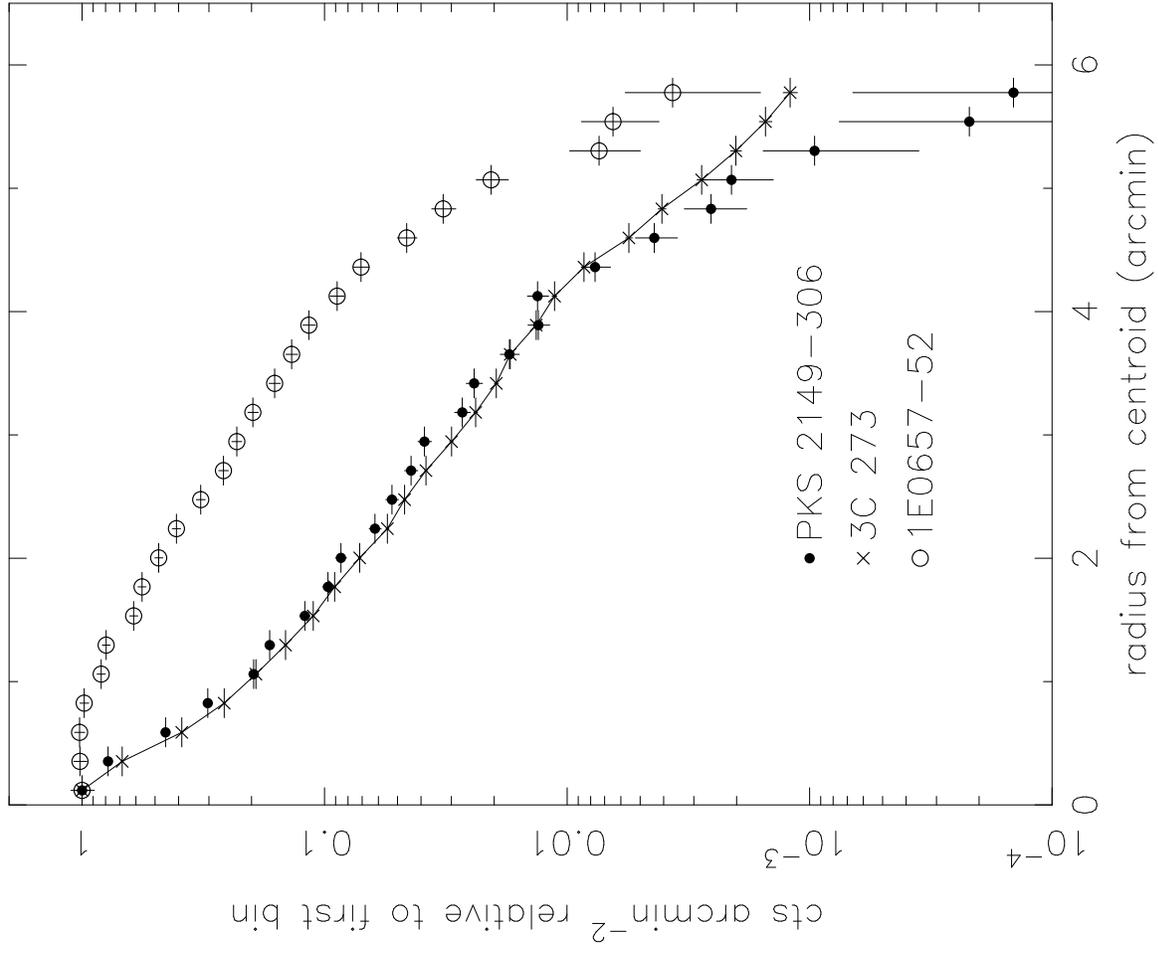
{\bf Figure 2} \\
The SIS0 Point Spread Function (PSF) for 
PKS 2149$-$306 (filled circles) compared with the
the PSF for 3C 273, a point source (crosses and solid line).
For comparison,
also shown is the PSF for the cluster of galaxies 1E0657$-$56 at $z=0.296$  
(open circles).
Note that the off-axis angles for all three sources are similar
so the PKS 2149$-$306 data are consistent with a point
source centered on the quasar (see text).

\par\noindent
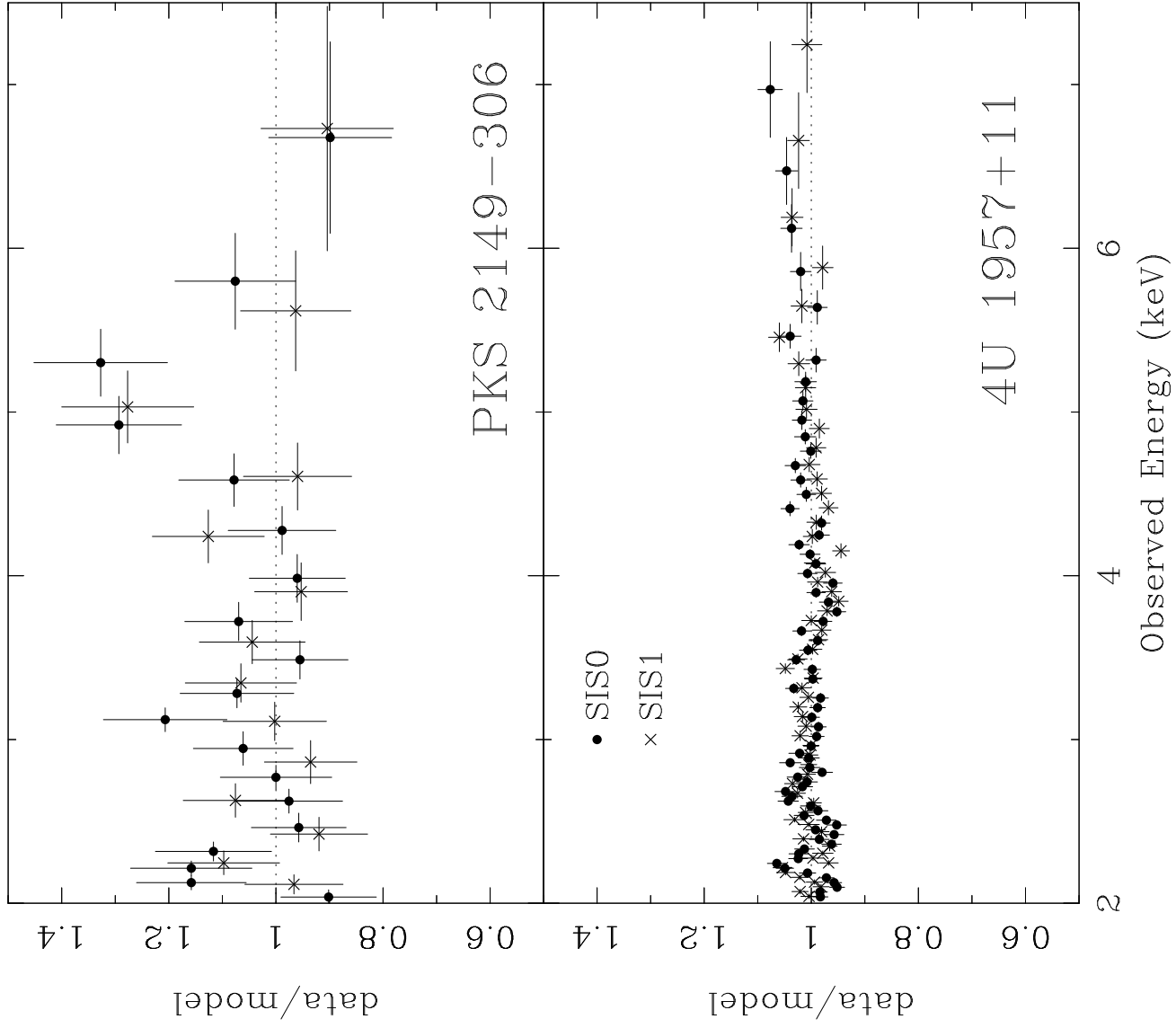
{\bf Figure 3} \\
Ratio of SIS spectral data to best-fitting 
power-law plus absorber models for 
4U~1957$+$11 and 
PKS 2149$-$306. The 4U~1957$+$11 data were taken just
four days after the PKS 2149$-$306 data.
Note that the vertical scale in both panels is identical.    
Thus, calibration uncertainties in the
SIS cannot account for the emission-line feature seen in 
PKS 2149$-$306 (also see text). 


\begin{references}

\reference{Ca97} Cappi, M., Matsuoka, M., Comastri, A., Brinkmann, W.,
		Elvis, M., Palumbo, G. C., \& Vignali, C. 1997, \apj, 478, 492 
\reference{dick1990} Dickey, J.M., \& Lockman, F.J. 
		1990, Ann. Rev. Ast. Astr. 28, 215. 
\reference{Fi98} Fiore, F., \etal 
		1998, \mnras, 298, 103  
\reference{3873-ovii} George, I.M., Turner, T.J., \& Netzer, H. 
		1995, \apjl, 438, L67
\reference{Nand94} Nandra, K., \& George, I.M. 1994, \mnras, 268, 405
\reference{Na97b} Nandra, K., George, I.M., Mushotzky, R.F., Turner, T.J.,
                 \& Yaqoob, T.
                   1997b, \apj, 476, 70
\reference{Nand96} Nandra, K., George, I.M., Turner, T.J., \& Fukazawa, Y.
	1996, \apj, 464, 165
\reference{Na97a} Nandra, K., Mushotzky, R.F., George, I.M., Turner, T.J.,
                \& Yaqoob, T.
                1997a, \apjl, 488, 91
\reference{Re97} Reeves, J.N., Turner, M.J.L., Ohashi, T., \& Kii, T.
		1997, \mnras, 292, 468 
\reference{Sieb1996} Siebert, J., Matsuoka, M., Brinkmann, W.,
        Cappi, M., Mihara, T., \& Takahashi, T. 1996, A\&A, 307, 8
\reference{Ta94} Tanaka, Y, Inoue, H., \& Holt, S.S. 
		1994, \pasj, 46, L37 
\reference{TonS180} Turner, T.J., George, I.M., Nandra, K. 
		1998, \apj, 508, 648 
\reference{Wilk86} Wilkes, B. J. 1986, \mnras, 218, 331
\reference{Yama97} Yamashita, A., Matsumoto, C., Ishida, M., Inoue, H,
	Kii, T., Makishima, K., Takahashi, T., \& Tashiro, M. 1997,
	\apj, 486, 763
\reference{Tazza98}     Yaqoob, T.
                1998, \apj, 500, 893
\reference{Tazza99}     Yaqoob, T.
                1999, \apj, 511, L75
\reference{Yaqo1998} Yaqoob, T., George, I. M., Turner, T. J.,
	Nandra, K., Ptak, A., \& Serlemitsos, P. J. 1998, \apj, 505, L87

\end{references}
\end{document}